\documentclass[12pt]{article}
 \usepackage{graphicx}
\usepackage{latexsym}
\usepackage{dcolumn}
\usepackage{stmaryrd}

\usepackage{graphicx}
\renewcommand{\baselinestretch}{1.8}

\begin{document}

\author{Attila R. Imre $^{(1)}$,
Alexander S. Abyzov $^{(2, 4)}$, Imre F. Barna $^{(1)}$, \\and  J\"urn W. P. Schmelzer $^{(3, 4)}$
\\[1ex]
 $^{(1)}$ KFKI Atomic Energy Research Institute,
\\H-1525 Budapest, POB 49, Hungary\\[1ex]
$^{(2)}$ National Science Center, \\Kharkov Institute of Physics and
Technology, Academician Str. 1, \\61108 Kharkov, Ukraine\\[1ex]
$^{(3)}$ Institut f\"ur  Physik der Universit\"{a}t Rostock,\\
Wismarsche Str. 43-45, 18051 Rostock, Germany\\
$^{(4)}$ Bogoliubov Laboratory of Theoretical Physics,\\ Joint Institute for Nuclear Research,  \\ul. Joliot-Curie 6, 141980 Dubna, Russia}

\title{{\bf Homogeneous bubble nucleation limit of mercury under the normal working conditions of the planned European Spallation Source}}
\date{}

\maketitle

\newpage

\begin{abstract}
\noindent In spallation neutron sources, liquid mercury is the subject of big thermal and pressure shocks, upon adsorbing the proton beam. These changes can cause unstable bubbles in the liquid, which  can damage the structural material. While there are methods to deal with the pressure shock, the local temperature shock cannot be avoided. In our paper we calculated the work of the critical cluster formation (i.e.  for mercury micro-bubbles) together with the rate of their formation (nucleation rate). It is shown that the homogeneous nucleation rates are very low even after adsorbing several proton pulses, therefore the probability of temperature induced homogeneous bubble nucleation is negligible.
\end{abstract}

\renewcommand{\baselinestretch}{2.0}

\section{Introduction}\label{sec.1}

Irradiating liquid metal (usually mercury) with proton beams is up to now  the best method to produce high-intensity, multi-purpose neutron beams. This method has been used in various existing facilities and it is planned to be used in the European Spallation Source, too. Unfortunately upon adsorbing the high-intensity proton beam in the liquid, the neutrons are not the only ones emitted; an unavoidable heat and pressure wave will be emitted simulataneously from the adsorption region. The increase of the temperature and (in the negative period of the pressure wave) the decrease of the pressure can cause cavitation in the liquid. The metal vapor bubbles then will flow with the liquid and upon reaching high pressure and low temperature regions, they will collapse, causing severe damage in nearby solid structures. This phenomenon is known as cavitation erosion and one of the main factors which (due to pitting and weight loss) shorten the lifetime of structural materials significantly. Therefore to avoid cavitation is one of the main challenges of the design of the spallation source target \cite{Liter.1, Liter.2, Liter.3}.

It should be mentioned here, the along the methods to minimize cavitation itself, there are two other ways to minimize the damage. One of them are the various ways of surface treatments (plasma nitriding, plasma carbonizing, etc.), which makes the surface more resistant to the damaging pressure wave emitted by the collapsing bubble \cite{Liter.4, Liter.5}. The other one is the addition of helium micro-bubbles, which is a proven way to soften up and to reduce the damage by absorbing the expansion of liquid mercury and mitigating the pressure waves \cite{Liter.4, Liter.5, Liter.6, Liter.7,  Liter.8}. Considering this method as a successful one to deal with the pressure-drop induced cavitation, in our paper we focused mainly on the temperature increase induced cavitation and allowed only small pressure changes to occur (down to -5 bar).

Our main aim is to determine whether the conditions (temperature and pressure changes) are able to cause cavitation or not. We approached the problem in three steps. In the first step (Section~\ref{sec.2}), we calculated the phase equilibrium, stability limit and various other properties of mercury by using the equation of state proposed by Morita et al. \cite{Liter.9,Liter.10, Liter.11, Liter.12}. In the next step (Section~\ref{sec.3}), we made some estimation for the magnitude of pressure and temperature changes by using single and repeated proton pulses on mercury. In the final step (Section~\ref{sec.4}), we calculated the work of critical bubble formation in mercury as well as the rate of homogeneous nucleation in the pressure-temperature range defined according to the results of the previous section. The paper is completed by a summary and discussion (Section~\ref{sec.5}).

\section{Model system}\label{sec.2}
\subsection{Location of binodal and spinodal curves}

For the description of mercury (Hg) in both the liquid and gas phases, we
will apply a slightly modified thermal equation of state as compared to the
expression proposed by Morita et al. (see \cite{Liter.9,Liter.10}, and, in
particular, Eq. (15) in \cite{Liter.11}). It reads
\begin{equation}  \label{Eq.1}
p=\frac{RT}{K(T)(v-b)}-\frac{a(T)}{v(v+c)}\;,
\end{equation}%
\begin{equation}  \label{Eq.2}
a(T)=a_{\mathrm{c}}\left( \frac{T}{T_{\mathrm{c}}}\right) ^{n}\qquad \mathrm{%
at}\qquad T\leq T_{\mathrm{c}}\;,
\end{equation}%
where $R=8.314\;\mathrm{J\cdot mol}^{\mathrm{-1}}$ is the universal gas
constant, $p$ is pressure, $v$ is molar volume, $T$ is temperature, $a_{%
\mathrm{c}}$, $b$, $c$ and $n$\ are the model parameters specific for the
substance considered, $T_{\mathrm{c}}$ \ is critical temperature. The
correction coefficient $K(T)$ is dependent on temperature only, it was
introduced in the repulsive term instead of the parameter $x_{\mathrm{d}}$,
which is a function of $T$\ and $p$ (see \cite{Liter.12}, in such case Eq.~(%
\ref{Eq.1}) becomes an equation for definition of $p(v,T)$, and has no
analytical solution).

We employ further dimensionless variables
\begin{equation}  \label{Eq.3}
\Pi =\frac{p}{p_{c}}\;,\qquad \omega =\frac{v}{v_{c}}\;,\qquad \theta =\frac{%
T}{T_{c}}\;,
\end{equation}%
where $v_{c}$ is the molar volume, $p_{c}$ the pressure both at the critical
point with the critical temperature, $T_{c}$. These parameters can be
determined from Eq.~(\ref{Eq.1}) in the common way via
\begin{equation}  \label{Eq.4}
\left( \frac{\partial p}{\partial v}\right) _{T}=\left( \frac{\partial ^{2}p%
}{\partial v^{2}}\right) _{T}=0\qquad \mbox{at}\qquad T=T_{c}\;.
\end{equation}%
The equation of state in reduced variables is given by
\begin{equation}  \label{Eq.5}
\Pi (\theta ,\omega )=\frac{\theta }{\chi _{c}(\theta )(\omega -\beta )}-%
\frac{\alpha (\theta )}{\omega (\omega +\delta )}\;.
\end{equation}%
Here
\begin{equation}  \label{Eq.6}
\chi _{c}(\theta )=\frac{p_{c}v_{c}}{RT_{c}}K(\theta )
\end{equation}%
is the reduced critical compressibility, and
\begin{equation}
K(\theta )=1.106697-0.106697\cdot \exp \left( \frac{\theta -1}{0.17026}%
\right) \;,  \label{Eq.7}
\end{equation}%
\begin{equation}
\alpha (\theta )=\frac{a_{\mathrm{c}}\theta ^{n}}{p_{c}v_{c}^{2}}=\alpha
\theta ^{n}\;,\qquad \beta =\frac{b}{v_{c}}\;,\qquad \xi =\frac{c}{v_{c}}\;.
\label{Eq.8}
\end{equation}%
According to \cite{Liter.11} we have then
\begin{equation}
\alpha =2.5272\;,\qquad \beta =0.3952\;,\qquad \xi =-0.16567\;,\qquad
n=-0.0284127\;.  \label{Eq.9}
\end{equation}%
From Eqs.~(\ref{Eq.1}) and (\ref{Eq.4}) we get \cite{Liter.11}
\begin{equation}  \label{Eq.10}
v_{c}=\;1.797\cdot 10^{-4}\;\mathrm{m}^{3}/\mathrm{kg}\;,\qquad \rho
_{c}=\;5566\;\mathrm{kg/m}^{3}\;,
\end{equation}
\[
p_{c}=\;158\cdot 10^{6}\;\mathrm{Pa}\;,\qquad T_{c}=1762\;\mathrm{K}\;.
\]

The location of the classical spinodal curve can be found via the
determination of the extrema of the thermal equation of state, $\Pi (\theta
,\omega )$ (Eq.~(\ref{Eq.5})) considering the temperature $\theta $ as
constant. By taking the derivative of $\Pi (\theta ,\omega )$ with respect
to $\omega $, we obtain from Eq.~(\ref{Eq.5}) the result
\begin{equation}
\frac{\partial }{\partial \omega }\Pi (\theta ,\omega )=\frac{\alpha (\theta
)(2\omega +\xi )}{\omega ^{2}(\omega +\xi )^{2}}-\frac{\theta }{\chi
_{c}(\omega -\beta )^{2}}=0\;.  \label{Eq.11}
\end{equation}%
For $\theta <1$, this equation has two positive solutions $\omega _{\mathrm{%
sp}}^{\mathrm{(left)}}$ and $\omega _{\mathrm{sp}}^{\mathrm{(right)}}$ for $%
\omega $ corresponding to the specific volumes of the both macrophases at
the spinodal curves (or at the limits of metastability).

Similarly, the binodal curves give for $\theta \leq 1$ the values of the
specific volumes of the liquid and the gas phases coexisting in thermal
equilibrium at a planar interface. From the left branch of the binodal
curve, we get the specific volume of the liquid phase ($\omega _{l}^{\mathrm{%
\ (eq)}}(\theta )=\omega _{\mathrm{b}}^{\mathrm{(left)}}(\theta )$), from
the right branch of the binodal curve, we obtain the specific volume of the
gas ($\omega _{g}^{\mathrm{(eq)}}(\theta )=\omega _{\mathrm{b}}^{ \mathrm{%
(right)} }(\theta )$). For $\theta =1$, both solutions coincide in the
critical point ($\omega _{l}^{\mathrm{(eq)} }=\omega _{g}^{\mathrm{(eq)}%
}=\omega _{c}=1)$, again. Consequently, in order to determine the specific
volumes of the liquid and the gas at some given temperature in the range $%
\theta \le 1$, we have to specify the location of the binodal curve.

The location of the binodal curve may be determined from the necessary
thermodynamic equilibrium conditions (for planar interfaces) -- equality of
pressure and chemical potentials -- via the solution of the set of equations
\begin{equation}
\Pi _{l}(\omega _{l},\theta )=\Pi _{g}(\omega _{g},\theta )\; ,\qquad \mu
_{l}(\omega _{l},\theta )=\mu _{g}(\omega _{g},\theta )\; .  \label{Eq.12}
\end{equation}
Here by $\mu $ the chemical potential of the atoms or molecules in the
liquid $(l)$ and the gas $(g)$ are denoted. Having at our disposal already
the equation for the reduced pressure (c.f. Eq.~(\ref{Eq.5})), we have now
to determine in addition the chemical potential in dependence on pressure
and temperature. This task will be performed in the next section.

\begin{figure}*
\scalebox{1.2}{
\rotatebox{0}{\includegraphics{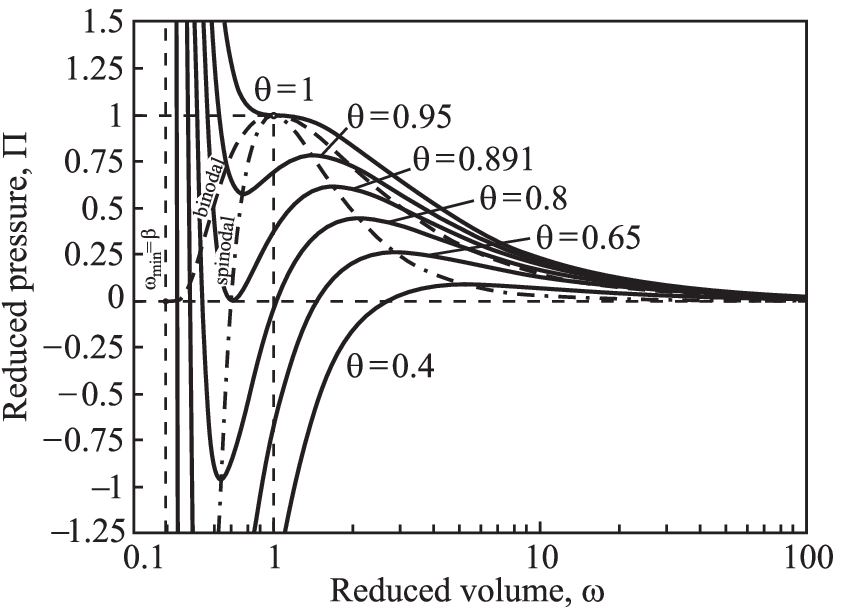}}}
\caption{  
Isotherms of mercury as described via Eq.(\ref{Eq.5}
) for different values of the reduced temperature.}	 
\label{figure.1}
\end{figure}

\begin{figure}*
\scalebox{1.2}{
\rotatebox{0}{\includegraphics{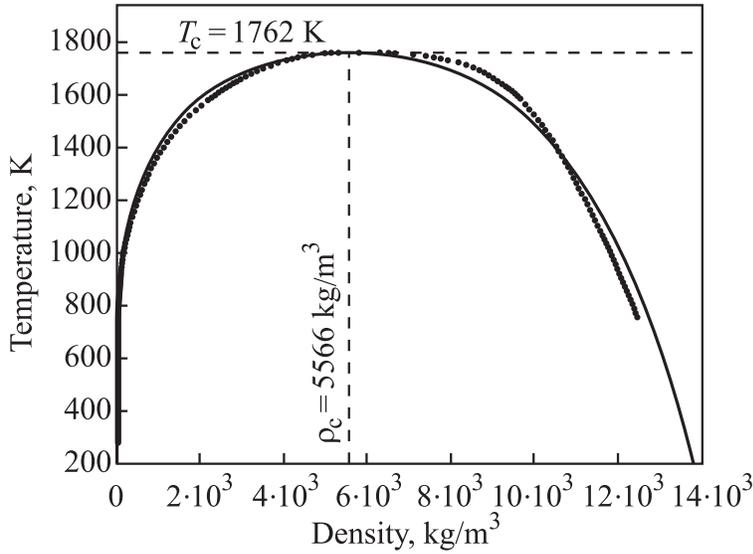}}}
\caption{  Comparison of experimental data (according to \cite{Liter.13,Liter.14}) for the vapor--liquid coexistence properties of mercury with the theoretical results (full curve determined via Eq.~(\ref{Eq.5})) obtained in this work.}
\label{figure.2} 
\end{figure}

\begin{figure}*
\scalebox{1.2}{
\rotatebox{0}{\includegraphics{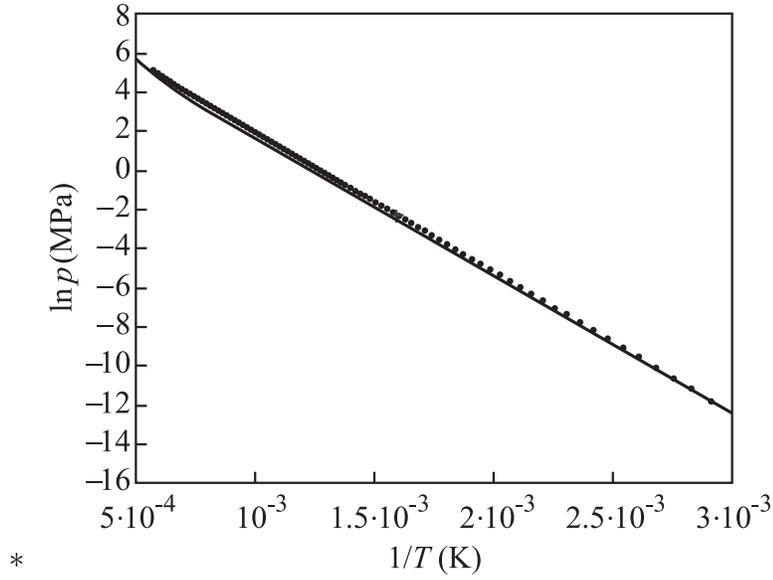}}}
\caption{ Comparison of experimental vapor pressure for mercury (according to \cite{Liter.13,Liter.14})
with theoretical results (full curve determined via Eq.(\ref{Eq.5})) obtained in this work.  
 }\label{figure.3}	 
\end{figure}

Isotherms for mercury according to Eq.~(\ref{Eq.5}) for different values of
the reduced temperature $\theta =0.4$, 0.65, 0.8, 0.891 and 0.92 are shown
in Fig.~\ref{figure.1}, dashed and dashed-dotted curves present binodal and
spinodal, correspondingly. One can see, that there are two classes of
isotherms: for the first one ($\theta \geq \theta _{s}$) $p\geq 0$, and for
the second class ($\theta <\theta _{s}$) pressure may be both positive and
negative.
The parameter $\theta _{s}$\ is determined via the equation%
\begin{equation}
\Pi _{l}(\omega _{\mathrm{sp}}(\theta _{s}),\theta _{s})=0,  \label{Eq.23}
\end{equation}%
for mercury $\theta _{s}\approx 0.891$ and $T_{s}\equiv T_{c}\theta
_{s}\approx 1570\;\mathrm{K}$.
A comparison of experimental data \cite{Liter.13, Liter.14} for the vapor--liquid
coexistence properties of mercury with results obtained in this work are
shown in Fig.~\ref{figure.2} and Fig.~\ref{figure.3} for ($T, \rho $) and ($
p,T^{-1}$)-variables, correspondingly.

\subsection{Determination of the chemical potential and the interfacial tension}

For processes at constant temperature, the change of the Helmholtz free
energy, $F$, may be expressed as
\begin{equation}
dF=-pdV+\mu dn\;.  \label{Eq.13}
\end{equation}%
Here $V$ is the volume of the system and $n$ the number of moles in it. For
a given fixed mole number, $n$, of the substance ($n=\mathrm{constant}$), we
have, in particular,
\begin{equation}
d\varphi _{n}=-pdv\;,\qquad \varphi _{n}=\frac{F}{n}\;,\qquad v=\frac{V}{n}%
\;,  \label{Eq.14}
\end{equation}%
or, in reduced variables,
\begin{equation}
d\left( \frac{\varphi _{n}}{p_{c}v_{c}}\right) =-\Pi d\omega \;.
\label{Eq.15}
\end{equation}%
Employing in the integration of Eq.~(\ref{Eq.15}) the equation of the state,
Eq.~(\ref{Eq.5}), we obtain
\begin{equation}
\left( \frac{\varphi _{n}}{p_{c}v_{c}}\right) =-\left[ \frac{\alpha (\theta )%
}{\xi }\ln \left( 1+\frac{\xi }{\omega }\right) +\frac{\theta }{\chi
_{c}(\theta )}\ln (\omega -\beta )\right] \;.  \label{Eq.16}
\end{equation}

Alternatively, the change of the Helmholtz free energy -- provided the
volume $V$ is fixed -- is given at constant temperature by
\begin{equation}
dF=\mu dn\;.  \label{Eq.17}
\end{equation}%
From Eq.~(\ref{Eq.17}), we arrive at
\begin{equation}
d\varphi _{v}=-\frac{\mu }{v^{2}}dv\;,\qquad \varphi _{v}=\frac{F}{V}\;.
\label{Eq.18}
\end{equation}%
On the other side, the functions $\varphi _{v}$ and $\varphi _{v}$ are
connected by
\begin{equation}
F=\varphi _{n}n=\varphi _{v}V\;,\qquad \varphi _{v}=\frac{\varphi _{n}}{v}\;.
\label{Eq.19}
\end{equation}%
With Eq.~(\ref{Eq.16}), we have then
\begin{equation}
\varphi _{v}=\frac{p_{c}}{\omega }\left[ \frac{\alpha (\theta )}{\xi }\ln
\left( 1+\frac{\xi }{\omega }\right) +\frac{\theta }{\chi _{c}(\theta )}\ln
(\omega -\beta )\right] \;.  \label{Eq.20}
\end{equation}%
With Eqs.~(\ref{Eq.18}) and (\ref{Eq.20}), the expression for the chemical
potential of a HLM can be obtained then via
\begin{equation}
\mu =-v^{2}\frac{\partial \varphi _{v}}{\partial v}=-v_{c}\omega ^{2}\frac{%
\partial \varphi _{v}}{\partial \omega }\;.  \label{Eq.21}
\end{equation}%
This relation yields
\begin{equation}
\frac{\mu }{p_{c}v_{c}}=-\left[ \frac{\alpha (\theta )}{\omega +\xi }+\frac{%
\theta \omega }{\chi _{c}(\theta )(\omega -\beta )}+\frac{\alpha (\theta )}{%
\xi }\ln \left( 1+\frac{\xi }{\omega }\right) +\frac{\theta }{\chi
_{c}(\theta )}\ln (\omega -\beta )\right] \;.  \label{Eq.22}
\end{equation}

\begin{figure}*
\scalebox{1.2}{
\rotatebox{0}{\includegraphics{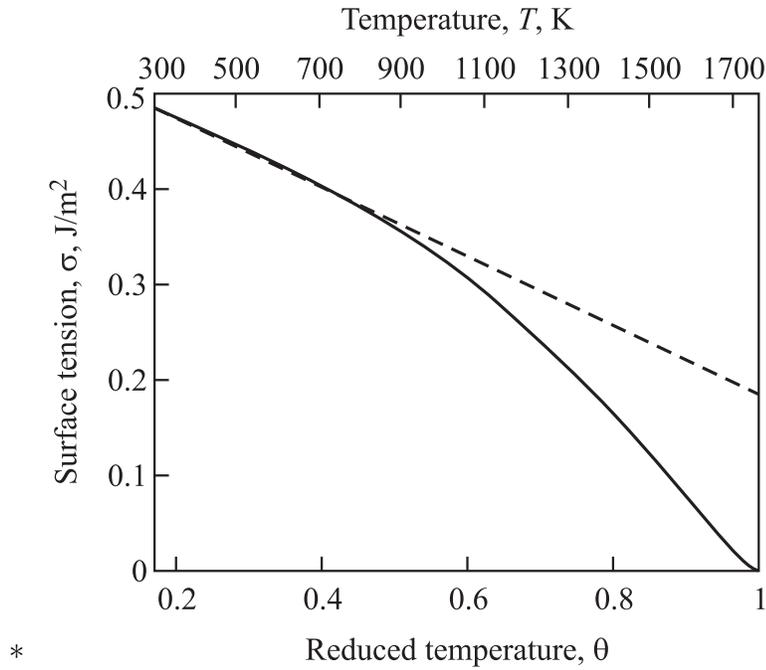}}}
\caption{ Dependence of the surface tension on temperature,
solid curve presents Eq.(\ref{Eq.24}) at $\omega _{\mathrm{l}}=\omega _{%
\mathrm{b}}^{\mathrm{(left)}}$, $\omega _{\mathrm{g}}=\omega _{\mathrm{b}}^{%
\mathrm{(right)}}$, and dashed curve -- Eq.(\ref{Eq.26}).}\label{figure.4}
\end{figure}

In addition to the bulk properties of the system under consideration, we
have to know the value $\sigma $ of the surface tension for a coexistence of
both phases at planar interfaces in dependence on the parameters describing
the state of both phases. We choose here this dependence in the form \cite%
{Liter.15, Liter.16, Liter.17, Liter.18}
\begin{equation}
\sigma \left( \omega _{\mathrm{g}},\omega _{\mathrm{l}},\theta \right)
=\Theta (\theta )\left[ \frac{1}{\omega _{\mathrm{l}}}-\frac{1}{\omega _{%
\mathrm{g}}}\right] ^{\delta }\;,\qquad \delta =2.5\;,  \label{Eq.24}
\end{equation}%
where%
\begin{equation}
\Theta (\theta )=A\left[ \frac{1}{\omega _{\mathrm{b}}^{\mathrm{(left)}}}-%
\frac{1}{\omega _{\mathrm{b}}^{\mathrm{(right)}}}\right] ^{n-\delta },
\label{Eq.25}
\end{equation}%
and $A$ and $n$ are constant parameters. Comparison of Eqs.~(\ref{Eq.24})
and (\ref{Eq.25}) with experimental data \cite{Liter.19}
\begin{equation}
\sigma (T)=0.5446544-0.000204917\cdot T  \label{Eq.26}
\end{equation}%
(valid in this form only for temperatures far below the critical temperature; here the temperature is given in Kelvin and the surface tension in J/m$^{2}$%
) at $\omega _{\mathrm{l}}=\omega _{\mathrm{b}}^{\mathrm{(left)}}$ and $%
\omega _{\mathrm{g}}=\omega _{\mathrm{b}}^{\mathrm{(right)}}$ yields%
\begin{equation}
A=0.033253\;\mathrm{J/m}^{2}\;,\qquad n=3\;.  \label{Eq.27}
\end{equation}
In Fig.~\ref{figure.4} dependence of the surface tension on temperature is
shown, solid curve presents Eq.~(\ref{Eq.24}) at $\omega _{\mathrm{l}%
}=\omega _{\mathrm{b}}^{\mathrm{(left)}}$, $\omega _{\mathrm{g}}=\omega _{%
\mathrm{b}}^{\mathrm{(right)}}$, and dashed curve -- Eq.~(\ref{Eq.26}).

\section{Determination of the pressure and temperature change after proton adsorption}\label{sec.3}

For the determination of the pressure and temperature change, a "one dimensional six-equation two-fluid model" was used, which is capable to describe transients like pressure waves, quick evaporation or condensation which is proportional to cavitation caused by energetic proton interaction in mercury target \cite{Liter.20}. The method was developed to describe the sudden and drastic steam condensation, called water hammer \cite{Liter.21, Liter.22}.

The model contains six first-order partial-differential equations which describe one-dimensional surface-averaged mass, momentum and energy conservation laws for both phases. A special numerical procedure ensures that shock-waves can be described without any numerical dispersion. With two major modifications this model can be applied to investigate the thermo-hydraulic properties of the planned mercury target in the European Spallation Source (ESS). These modifications are the following: the equation of state namely the density and the internal energy of both mercury phases should be known in a broad range of pressure (1 Pa to 100 MPa) and temperature (273 K to 1000 K).
As a second point the interaction of the high energy proton beam with  mercury has to be included. This is a much simpler task because we may consider that about 50 \% of the 300 kJ/pulse beam energy is absorbed as a 2ms long heat shock square pulse, giving a new source terms in the energy equation of the liquid phase. The ESS mercury target station is modeled as  a 18 meter long closed loop which is in three dimension the pipe diameter 15 cm. We consider that 150 kJ heat is absorbed in a 10 cm long pipe, this is approximately the width of the proton pulse. Calculation shows that such a single pulse heats up the mercury with about 40-44 K, assuming that the initial temperature was between 293-373 K (i.e. within the normal working range of the spallation source).
 In the calculations, low velocities 0.5-4 m/s, low initial pressure 1-4 bar and low initial temperature (below 374 K) were assumed. To our knowledge the existing Japanese Spallation Neutron Source Hg loop is about 15 m long, with a diameter of 15 cm, the flow velocity of Hg is 0.7 m/s and the pressure is approximately equal to 1 bar.

Concerning the pressure change, the model is able to estimate the positive part, but at the negative region (where most of the low temperature cavitation is expected to  happen \cite{Liter.23}), a stability problem aroused. Therefore we focused our calculation to the heat shock and, at present, neglected the pressure change. Preliminary calculation yielded a few bar changes \cite{Liter.20}, in agreement with the results of Ida \cite{Liter.24, Liter.25}, therefore the latter calculations were performed in the -5 to 10 bar range. We should mention here, that other models predicted much larger pressure changes (even hundreds of bars) \cite{Liter.26, Liter.27} both in the positive and negative pressure region.

Also the effects of repeated pulses were checked.  The calculations were performed with a 2 ms square pulse train where the delay time was 20 ms which is similar to a 16Hz repetition rate. We started with a flow system with initial p=4 bar, $T_{initial} = 353$ K and initial flow velocity v = 4 m/s.  We found that the temperature jumps are more or less additive which means that after the beginning of the third pulse the temperature was about 430 K.

\section{Determination of the work of critical cluster formation}\label{sec.4}

Let us assume, now, that the system is brought suddenly into a metastable
state located between binodal curve and spinodal curve at the liquid branch
of the equation of state. Then, by nucleation and growth processes, bubbles
may appear spontaneously in the liquid and a phase separation takes place \cite{Liter.28}.
Based on the relations outlined above, we will determine now the parameters
of the critical clusters governing bubble nucleation in dependence on the
state parameters, pressure and temperature.

\begin{figure}*
\scalebox{1.2}{
\rotatebox{0}{\includegraphics{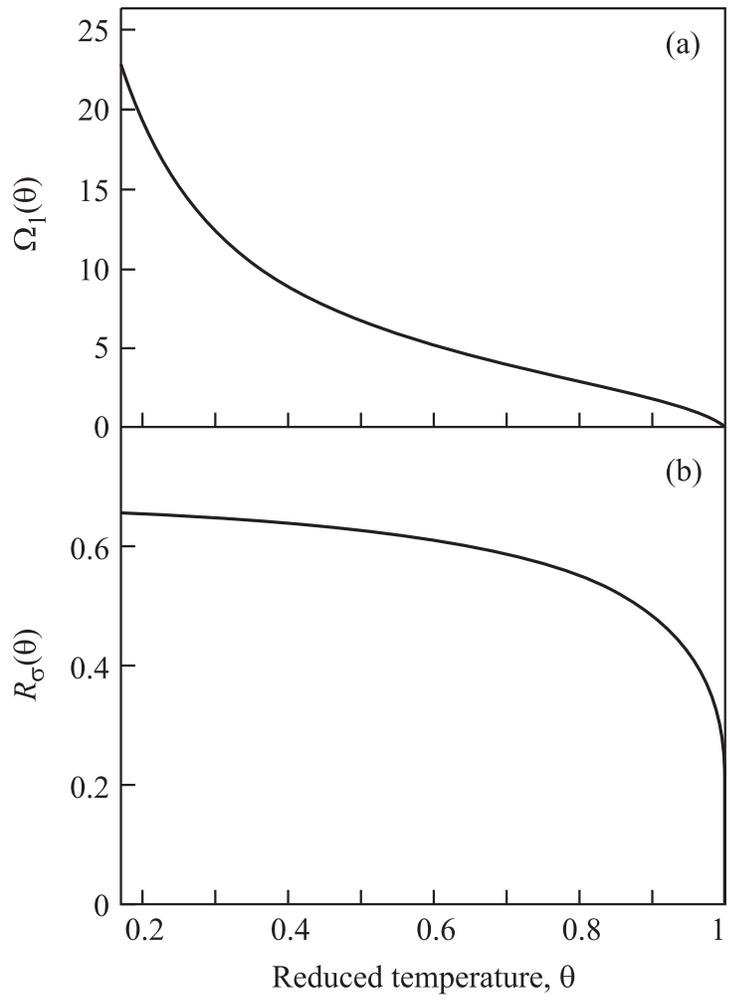}}}
\caption{ Dependence of the scaling parameters $\Omega _{1}$
and $R_{\sigma }$ on the reduced temperature, $\theta $.}
\label{figure.5}
\end{figure}

We start with the general expression for the change of the thermodynamic
potential
\begin{eqnarray}\label{Eq.3.4}
        \Delta G =
        \sigma A + (p-p_{\alpha})V_{\alpha}
        + \sum \limits_j n_{j\alpha}\left[
        \mu_{j\alpha}-\mu_{j\beta}\right]\; .
\end{eqnarray}
Here the subscript $\alpha$ specifies the parameters of the cluster (bubble) phase while $\beta$ refers to the ambient liquid phase. This relation holds generally provided - as we assume - the state of the ambient liquid phase remains unchanged by the formation of one bubble.
For a one-component system (as discussed here), this expression is reduced to
\begin{eqnarray}\label{Eq.3.4a}
        \Delta G =
        \sigma A + (p-p_{\alpha})V_{\alpha}
        + n_{\alpha}\left[
        \mu_{\alpha}-\mu_{\beta}\right]\; .
\end{eqnarray}
As the independent variables, we select the size of the bubble, $r$ and the molar volume of the gas phase in the bubble.
  Similarly to  \cite{Liter.15, Liter.29,Liter.30}, we arrive then at
\begin{equation}
\frac{\Delta g(r,\omega _{\mathrm{g}},\omega _{\mathrm{l}},\theta )}{k_{%
\mathrm{B}}T}=3\left( \frac{1}{\omega _{\mathrm{l}}}-\frac{1}{\omega _{%
\mathrm{g}}}\right) ^{\delta }r^{2}+2f\left( \omega _{\mathrm{g}},\omega _{%
\mathrm{l}},\theta \right) r^{3}\;,  \label{Eq.28}
\end{equation}
where the following notations have been introduced:
\begin{equation}
f\left( \omega _{\mathrm{g}},\omega _{\mathrm{l}},\theta \right) =\Pi
(\omega _{\mathrm{g}},\theta )-\Pi (\omega _{\mathrm{l}},\theta )+\frac{1}{%
\omega _{\mathrm{g}}}\left( \frac{\mu \left( \omega _{\mathrm{l}},\theta
\right) -\mu \left( \omega _{\mathrm{g}},\theta \right) }{p_{c}v_{c}}\right)
\;,  \label{Eq.29}
\end{equation}%
\begin{equation}
g\equiv \frac{G}{\Omega _{1}}\;,\qquad \Omega _{1}=\frac{16\pi }{3}\frac{1}{%
p_{c}^{2}k_{\mathrm{B}}T_{c}\theta }\Theta (\theta )^{3}\;,  \label{Eq.30}
\end{equation}%
\begin{equation}
r\equiv \frac{R}{R_{\sigma }}\;,\qquad R_{\sigma }=\frac{2}{p_{c}}\Theta
(\theta )\;.  \label{Eq.31}
\end{equation}%
The dependence of the scaling parameters $\Omega _{1}$ and $R_{\sigma }$ on
the reduced temperature is shown in Fig.~\ref{figure.5}.

\begin{figure}*
\scalebox{1.2}{
\rotatebox{0}{\includegraphics{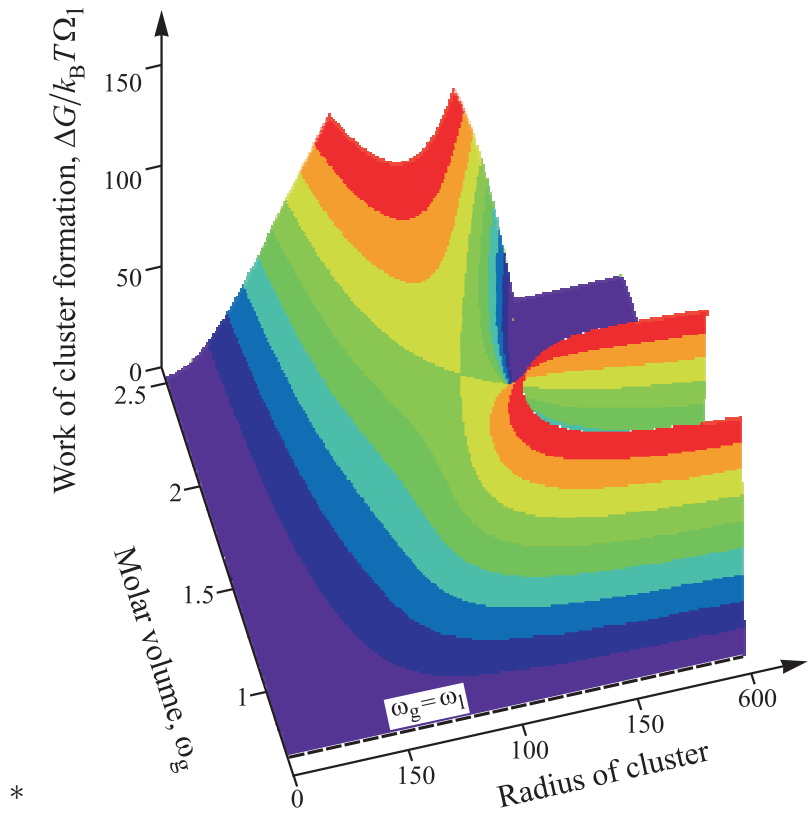}}}
\caption{  Gibbs free energy surface for metastable initial
state, $\theta =0.92$, $\omega _{\mathrm{l}}=0.65$.}
\label{figure.6}
\end{figure}

The Gibbs free energy surface for the metastable initial state has typical
saddle shape at the critical point (see Fig.~\ref{figure.6}, $\theta =0.92$,
$\omega _{\mathrm{l}}=0.65$). The critical point position is determined by
the set of equations%
\begin{equation}
\frac{\partial \Delta g(r,\omega _{\mathrm{g}},\omega _{\mathrm{l}},\theta )%
}{\partial r}=0\;,\qquad \frac{\partial \Delta g(r,\omega _{\mathrm{g}%
},\omega _{\mathrm{l}},\theta )}{\partial \omega _{\mathrm{g}}}=0\;.
\label{Eq.32}
\end{equation}

\begin{figure}*
\scalebox{1.2}{
\rotatebox{0}{\includegraphics{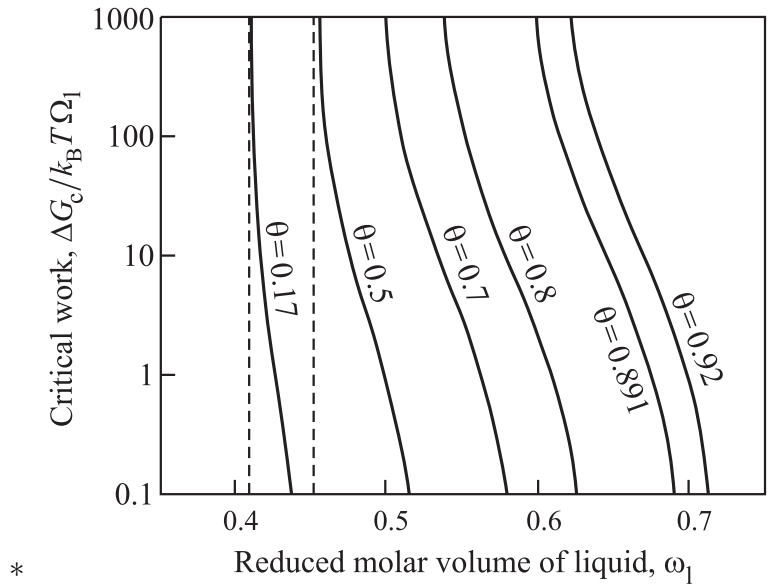}}}
\caption{   Dependence of the critical cluster radius, $%
r_{c}=R_{c}/R_{\sigma }$, on the initial molar volume of liquid, $\omega _{%
\mathrm{l}}$, for different values of temperature, $\theta =0.17$,\ 0.5,
0.7,\ 0.8,\ 0.891 and 0.92.}
\label{figure.7}
\end{figure}

The dependence of the critical cluster parameters on the initial molar volume of
liquid, $\omega _{\mathrm{l}}$, are shown in Figs.~\ref{figure.7}--\ref%
{figure.9}, for different values of temperature, $\theta =0.17$,\ 0.5, 0.7,\
0.8,\ 0.891 and 0.92. The positions of the binodal curves are given then by $
\omega _{\mathrm{b}}^{\mathrm{(left)}}=0.409,$\ 0.45, 0.494, 0.531, 0.589,
0.62, and $\omega _{\mathrm{b}}^{\mathrm{(right)}}=1.663\cdot 10^{8},$\
90.5, 11.606, 5.634, 3.043, 2.475, the respective parts of the spinodal
curves are located at $\omega _{\mathrm{sp}}^{\mathrm{(left)}}=0.452,$
0.528, 0.59, 0.634, 0.696, 0.726, and $\omega _{\mathrm{sp}}^{\mathrm{(right)%
}}=13.609,$\ 4.04, 2.584, 2.087, 1.679, 1.547, correspondingly.

\begin{figure}*
\scalebox{1.2}{
\rotatebox{0}{\includegraphics{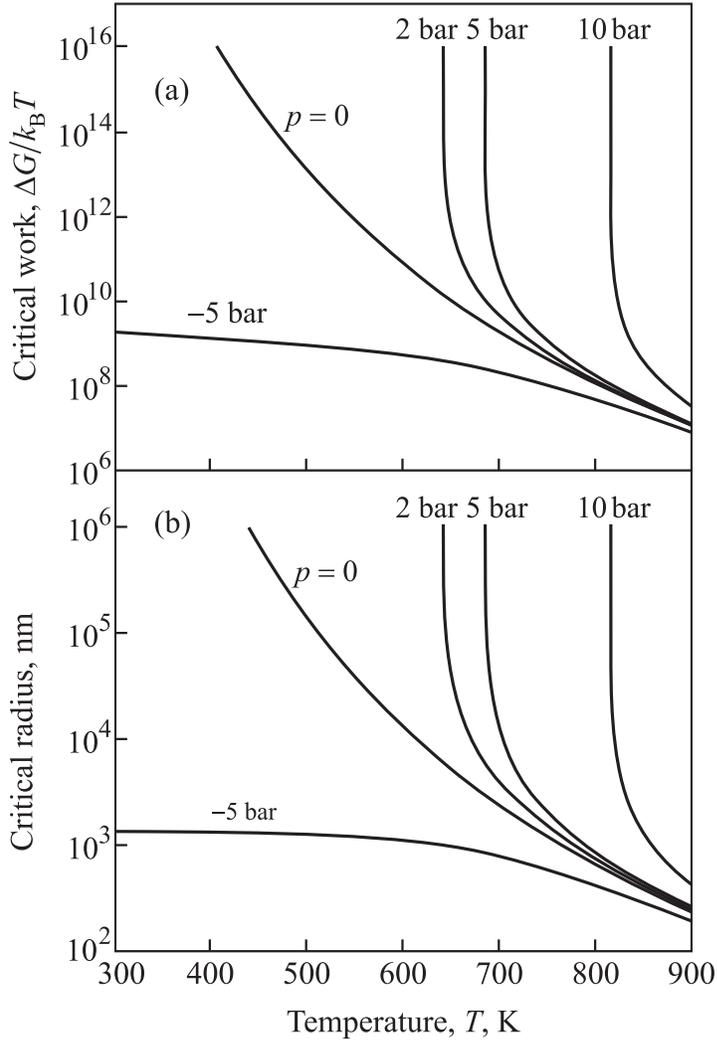}}}
\caption{  Dependence of the work of critical cluster
formation, $\Delta G_{c}/k_{B}T$ (a), and of the critical cluster radius (b)
on temperature
for $p=-5$, 0, 1, 2, 5 and 10 bar. }
\label{figure.10}
\end{figure}

\begin{figure}*
\scalebox{1.2}{
\rotatebox{0}{\includegraphics{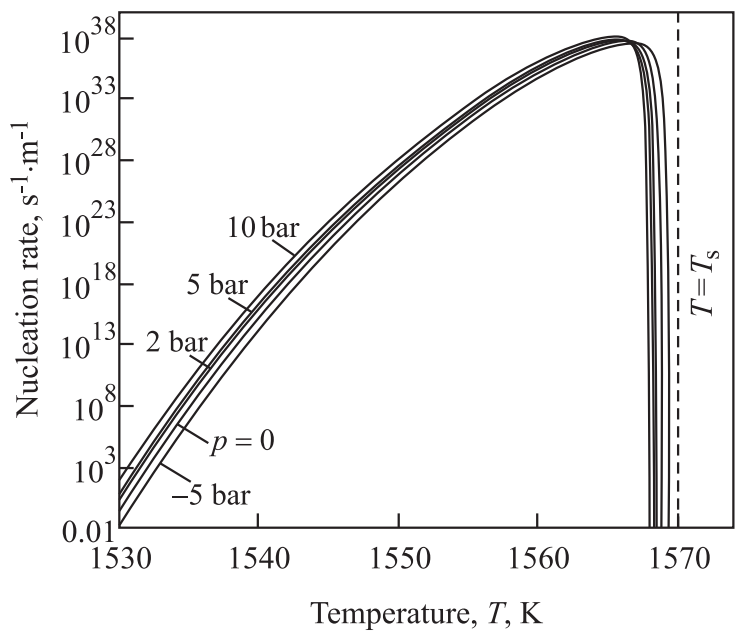}}}
\caption{  Dependence of the nucleation rate on temperature
for $p=-5$, 0, 1, 2, 5 and 10 bar. }
\label{figure.11}
\end{figure}

The dependence of the work of formation and radius of the critical cluster
on temperature for the practically significant cases $p=-5, 0, 1, 2, 5$ and $10$ bar
is presented in Fig.~\ref{figure.10}. In Fig.~\ref{figure.11}, the
dependence of the nucleation rates on temperature
for the same values of pressure are shown (a value of the
pre-exponential factor of $J_0=10^{41}$s$^{-1}$m$^{-3}$  has been used for the calculations).
One can see, that in such case homogeneous nucleation is possible only at
very high temperatures, near $T_{s}\approx 1570\;\mathrm{K}$. One can observed as well  that concerning a 20 cm diameter sphere (region of proton adsorption) and 2 ms time span, one can expect 1 or more nucleation event above 1530.5 K. However, heterogeneous nucleation may occur also at lower temperatures.

\section{Conclusions}\label{sec.5}

In spallation neutron sources, liquid mercury is the subject of big thermal and pressure shocks (including negative part), upon adsorbing the proton beam. Increased temperature and decreased pressure can cause instable bubbles which can cause cavitation erosion of the structural material, shortening the life-time of the equipment and contaminating the mercury with tiny steel pieces. Therefore it is crucial to avoid or minimize bubble nucleation. While pressure shock can be softened by adding helium micro-bubbles to the mercury, there is no way to deal with the thermal shock (i.e. local heating is not possible in the middle of the liquid mercury). Therefore our calculation focused on to calculate the extent of the temperature increase, the work of critical cluster formation (i.e. the nucleus of a macroscopic bubble) and the nucleation rate. It has been shown that after repeated proton pulses the temperature can be increased with a few hundred K, but the nucleation rate is so low that the possibility of homogeneous nucleation (i.e. bubble formation in the pure mercury) is highly improbable, even when the pressure goes below the vapor pressure.

\section{Acknowledgments}

The authors express their gratitude to the Hungarian Academy of Sciences for
financial support in the framework of the common Hungary-Dubna project EAI-2009/004.

\newpage 

\pagebreak

\pagebreak

\pagebreak

\pagebreak 

\pagebreak 

\pagebreak

\pagebreak 

\begin{figure}*
\scalebox{1.2}{
\rotatebox{0}{\includegraphics{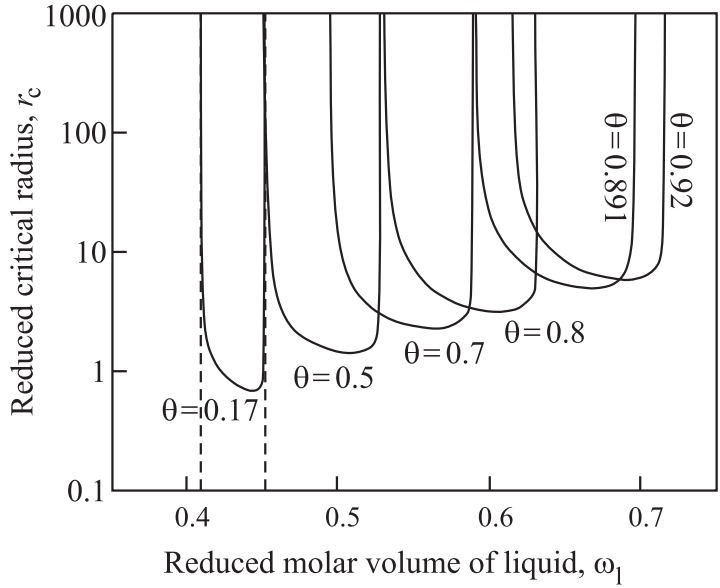}}}
\caption{   Dependence of the gas molar volume in critical
bubble, $\omega _{\mathrm{g,c}}$, on the initial molar volume of liquid, $%
\omega _{\mathrm{l}}$, for different values of temperature, $\theta =0.17$,\
0.5, 0.7,\ 0.8,\ 0.891 and 0.92.
 }\label{figure.8}
\end{figure}

\pagebreak

\begin{figure}*
\scalebox{1.2}{
\rotatebox{0}{\includegraphics{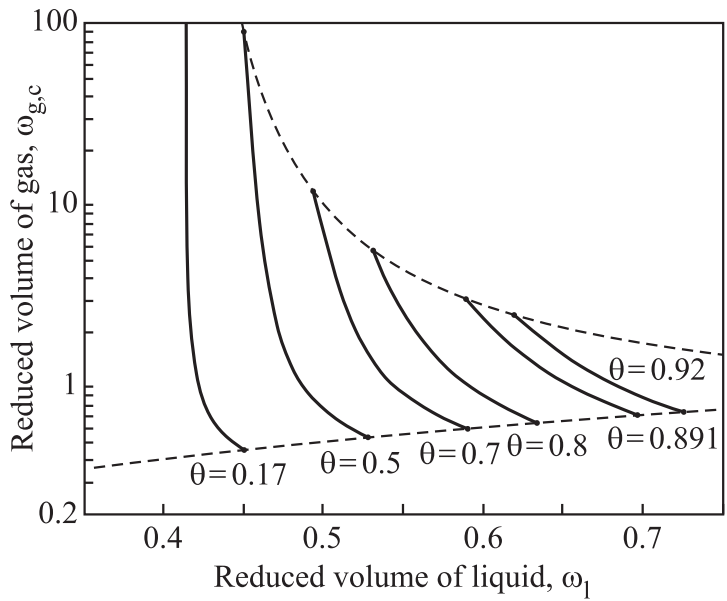}}}
\caption{   Dependence of the work of critical cluster formation,
$\Delta G_{c}/k_{B}T\Omega _{1}$, on the initial molar volume of liquid, $%
\omega _{\mathrm{l}}$, for different values of temperature, $\theta =0.17$,\
0.5, 0.7,\ 0.8,\ 0.891 and 0.92. }
\label{figure.9}
\end{figure}

\pagebreak 

\pagebreak

\end{document}